\documentclass[aps,prl,preprint,showpacs]{revtex4}
\usepackage{graphicx}
\usepackage{amssymb}
\usepackage{amsfonts}
\usepackage{amsmath}
\usepackage{dcolumn}
\usepackage{natbib}
\usepackage{color}

\begin{document}
\title{First Principles Study of Bismuth Films on the Nickel(111) Surface}
\author{Qin Gao and Michael Widom}
\affiliation{Department of Physics, Carnegie Mellon University, Pittsburgh, PA 15213}

\begin{abstract}
A recent experiment~\cite{Bollmann11} suggested that bismuth forms hexagonal close packed (HCP) films on the Ni(111) surface, of heights 3, 5 and 7 layers. A quantum size effect based on free electrons was proposed in explanation. To test this idea, we calculate the total energies of Bi on the Ni(111) surface using density functional theory. We find that HCP film stabilities disagree with the observed odd layer preferences, and the structures are mechanically destabilized by adding capping atoms which pucker the HCP layers. Furthermore, we find that rhombohedral films based on the bulk Bi structure are energetically more favorable than the proposed HCP films. These structures also favor odd numbers of layers, but owing to covalent chemical bonding rather than confinement of free electrons. Specifically, a strongly bound adsorbed surface monolayer forms, followed by bulk-like rhombohedral bilayers.

\end{abstract}

\maketitle

\section{Introduction}

\indent During the growth of thin metallic films, confinement of electrons can favor film heights that are commensurate with half the Fermi wavelength~\cite{Zhang98}. This well known quantum size effect due to confinement (QSE) has long been studied in various materials, including bismuth.  Bismuth differs from conventional QSE elements because it is a semimetal with an especially long Fermi wavelength ($\sim$ 40 nm)~\cite{Hofmann06,Edelman76} that causes various physical properties of Bi films to oscillate with long periods~\cite{Rogacheva03}. In ultrathin films ($\sim$ 1 nm), Bi exhibits allotropic transformations from puckered pseudocubic films to hexagonal symmetry films on Si(111) and on some quasicrystal surfaces~\cite{Nagao04,Fournee05}. Experiments and first principles calculations concur that both film types exhibit bilayer growth, due to the exotic bonding character of Bi, rather than QSE~\cite{Yaginuma07}. In contrast, the initial growth of Bi films on metallic substrates has not been well studied until a recent experiment~\cite{Bollmann11} reported stable 3, 5 and 7 layer Bi hexagonal close packed (HCP) films of Bi on the Ni(111) surface. They attempted to explain the stability by quantum confinement based on a free electron model. A more in-depth theoretical study is needed to understand this possible short period QSE in Bi.

Bulk Bi takes the rhombohedral structure of Pearson type hR2 (prototype $\alpha$-As) common to group-V semimetals, which is distorted from the simple cubic structure by a Jones-Peierls mechanism~\cite{Jones34}.  The bulk Bi structure is best described as a stacking of bilayers in the [001] direction. Here we use 3-element hexagonal indexing, it would be [111] using rhombohedral indexing. Each bilayer has height 1.43~\AA~ and is separated from the adjacent bilayer by 2.51~\AA.  Within the (001) plane the Bi spacing is 4.53~\AA.  However, three strong covalent bonds of length 2.98~\AA~ link each Bi atom to others within each bilayer, while three weak metallic bonds of length 3.62~\AA~ connect each Bi atom to others in the adjacent bilayer.  The (001) plane is thus a natural cleavage plane, with divisions expected between bilayers.  In addition to hR2, bulk Bi possesses many allotropes, especially at high pressure, including Pearson structure types mP4, mC4, cP1 and cI2, but it does not take the HCP structure.

In their experiment~\cite{Bollmann11}, the authors grew Bi on a Ni(111) surface at a temperature of 473 K via vapor deposition.  Based on the low energy electron diffraction (LEED) patterns, deposition and film growth rates, and low energy electron reflectivity spectra, the authors proposed that initially a (7$\times$7) wetting layer forms (hereafter referred to as adsorbed surface monolayer), which transitions to a 3 layer HCP film with a (3$\times$3) surface cell sitting directly on the Ni substrate, as well as a 5 layer HCP film with a [3-112] surface cell, as coverage grows. However, at 422 K, Bi formed a 7 layer film with an (8$\times$8) cell~\cite{Bollmann12} surrounded by the 3 layer (3$\times$3) film. The proposed 3 layer and 5 layer Bi films on Ni have in-plane lattice constant 3.7-3.8~\AA~(See Appendix A1. for  discussion of lateral strain).  Taking their measured lattice constants and an assumed free electron valence of 5, they calculated that their 3, 5 and 7-layer films were, respectively, 2.5, 4.0 and 5.0 Fermi wavelengths in height. As these structures and bond lengths have not been previously observed, and Bi is notoriously {\em not} free electron-like, a first principles electronic structure investigation is warranted.

\section{Methods}

\indent We apply electronic density functional theory, using the Vienna ab-initio simulation package (VASP~\cite{Kresse93,Kresse96}) to solve the Kohn-Sham equations with the Perdew-Burke-Ernzerhof (PBE~\cite{Perdew96}) parameterization of the generalized gradient approximation (GGA) for the electron exchange correlation potential. We use projector augmented wave potentials~\cite{Blochl94,Kresse99} with a fixed energy cutoff of 269.5 eV (the default for Ni). The d semi-core levels of Bismuth are treated as valence electrons. Collinear spin polarization is used since Ni is ferromagnetic, though we test noncollinear magnetism to check the importance of spin-orbit coupling (SOC) for some structures. In the noncollinear calculations, we take the relaxed structure from collinear calculations and perform a static calculation.

We construct models based on four Ni layers normal to the (111) surface with Bi films on one side. Our cells include 22 ~\AA~of vacuum, with periodic boundary conditions.  Electrostatic energy created by the asymmetric charge distribution in the presence of Bi is small relative to the differences of surface energies.  All structures are relaxed holding the cell sizes and bottom layer Ni atoms fixed, with in-plane lattice constants set by the relaxed bulk Ni structure. Energy convergence is carefully checked with respect to the vacuum size, k-point mesh and the number of Ni layers.

\section{Results and Discussion}

\begin{figure}[ht!]
\includegraphics[scale=0.45,width=0.5\textwidth,clip]{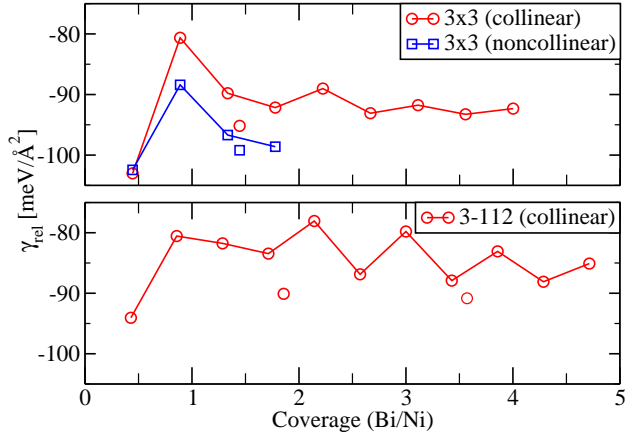}
\caption{(Color online) Relative surface energies of HCP Bi films with (3$\times$3) and [3-112] Ni (111) surface cells. Note the energy conversion factor is 1 eV/\AA$^2$=16.0 J/m$^2$. Red points are from collinear calculations, blue points are noncollinear. Spin-orbit coupling affects the total energy quantitatively but does not alter the relative film stabilities. Data points connected by line segments correspond to coverages of integer numbers of monolayers (1-9 for 3$\times$3, 1-11 for [3-112]).  Extra data points in the (3$\times$3) cell are three layer films with one extra capping atom at a valley site. Extra points in the [3-112] cell are 4 layer plus one atom, and 8 layer plus one atom at valley sites.}
\label{fig:grel}
\end{figure}

We first compare the relative surface energies of the proposed Bi HCP films of different thickness on Ni(111). Several quantities are needed to define relative surface energy: the total energy $E_{tot}$ of the $N_{Bi}$ atoms of Bi on the surface of the 4-layer Ni slab; the slope $E_{Bi}$, which is the linear part of $E_{tot}$ as it depends on $N_{Bi}$; the energy $E_{Ni}^{slab}$ of the Ni slab including its two free surfaces, each of area A.  The slope $E_{Bi}$ can be considered as the energy of bulk Bi in the HCP structure with the in-plane lattice constant determined by the surface cell~\cite{Fiorentini96}. With these definitions, relative surface energy is
\begin{equation}
\label{eq:surfNi}
\gamma_{rel}=[E_{tot}-E_{Ni}^{slab}-E_{Bi}N_{Bi}]/A.
\end{equation}
 Fig.~\ref{fig:grel} shows the relative surface energies of (3$\times$3) and [3-112] HCP films. A structure is relatively stable if the second derivative of its relative surface energy with respect to film thickness is positive~\cite{Wu08}. Relative stability occurs for (3$\times$3) cells of 1, 3, 4, 6 and 8 layers and for [3-112] cells of 1, 4, 6, 8 and 10 layers. These predicted stabilities disagree with the experimentally discovered odd layer preference. 

Also shown in Fig.~\ref{fig:grel} are data points for films with single Bi adatoms.  These data points lie below the relative surface energies of integer layer coverages, revealing that terminating on complete HCP layers is unfavorable.  Strong puckering of the HCP layers occurs during relaxation of structures with adatoms.  Relative surface energies calculated with SOC are shown for the (3$\times$3) films. SOC influences the relative surface energies quantitatively but does not alter the sequence of stable structures.

\begin{figure}[ht]
\includegraphics[scale=0.4,clip]{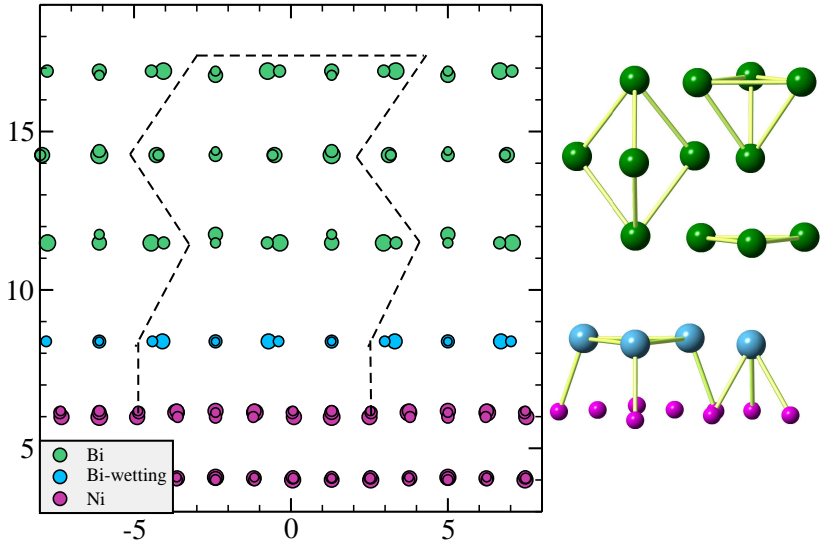}
\caption{(Color online) Side view of relaxed relatively stable 4-layer HCP Bi film on Ni(111) (3$\times$3) cell (units are ~\AA). Chemical bonding shown at right (Slightly tilted).}
\label{fig:HCP4}
\end{figure}

The relatively stable 4-layer (3$\times$3) structure is illustrated in Fig.~\ref{fig:HCP4}. The first layer Bi atoms strongly bond with the Ni surface atoms. On top of the first layer, Bi atoms form slightly puckered layers in order to achieve short Bi bonds with adjacent layers.  These short covalent bonds lower the relative surface energy. 

The asymptotic period 2 oscillations of relative surface energies are reminiscent of the QSE prediction, but the minima occur for even numbers of HCP layers rather than odd. Moreover, we have done a genuine first principles study of QSE in free standing HCP Bi films (see Appendix A2.) and find that the actual predicted oscillation period is close to 3 layers, not 2. Thus our first principles calculations cast doubt on both the HCP structure model and the proposed explanation in terms of QSE.

\begin{figure}[ht]
\includegraphics[scale=0.4]{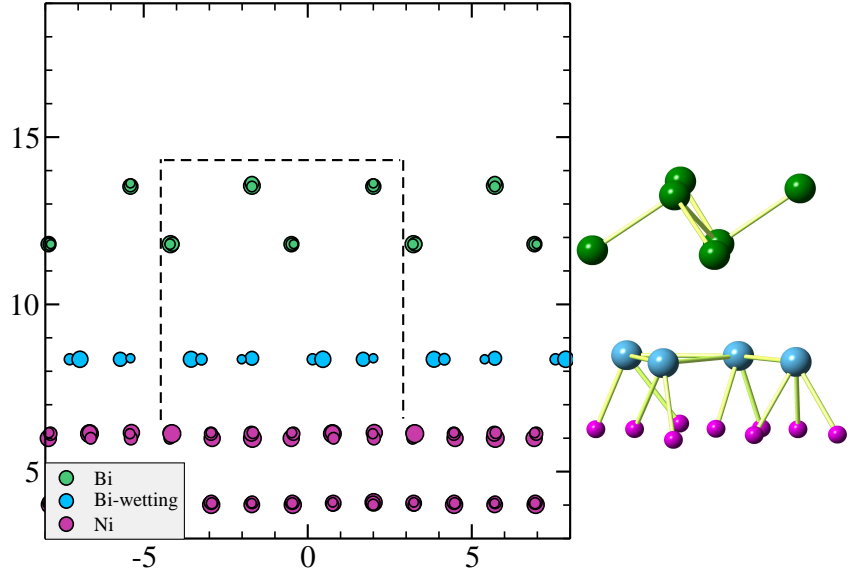}
\caption{(Color online) Side view of relaxed Bi surface monolayer and rhombohedral bilayer on Ni(111) (3$\times$3) cell (units are ~\AA). Chemical bonding shown at right (Slightly tilted).}
\label{fig:hR2}
\end{figure}

We now seek alternate film structures. Inspecting Fig.~\ref{fig:HCP4} we note the interaction of the upper layer Bi with the surface monolayer is weak, so strong deviations of structure and bonding from bulk Bi due to the Ni substrate are not anticipated. Thus bulk-like hR2 Bi (001) films are good candidates to grow on top of the surface monolayer. Henceforth, when we say ``hR2 film'' we include the surface monolayer.  For example, a 3 layer hR2 film (see Fig.~\ref{fig:hR2}) consists of the surface monolayer plus an hR2 (001) bilayer. The in-plane lattice constant of bulk hR2 Bi is 4.53~\AA (see Appendix A1.). Meanwhile, $\sqrt{3}$ times the Ni interatomic spacing is 4.31~\AA~which differs by only 5\%.  Furthermore, in our calculated energy of strained free standing Bi (001) films, the energy minimum occurs at in plane lattice constants 4.3~\AA, 4.4~\AA~and 4.5~\AA~for one, two and three bilayer films respectively. Thus the stable thin bilayer films match with Ni very well. 

To illustrate the relative stability of various structures, we compare the surface enthalpy of formation, which is defined as,
\begin{equation} 
\label{eq:enthalpy}
\Delta H/A=[E_{tot}-E_{Ni}^{slab}-E_{Bi}^{bulk}N_{Bi}]/A
\end{equation}
which differs from the relative surface energy $\gamma_{rel}$ in Eq.~\ref{eq:surfNi} only in our choice of reference energy for pure Bi, $E_{Bi}^{bulk}$ is the relaxed Bi bulk energy in the hR2 structure.  

\begin{figure}[ht]
\includegraphics[scale=0.45,width=0.5\textwidth,clip]{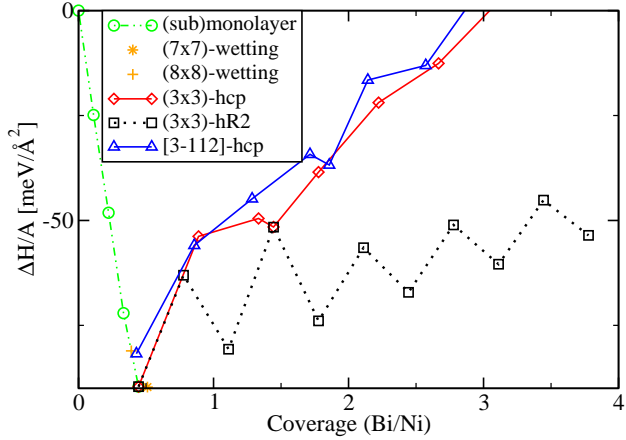}
\caption{(Color online) Enthalpy of formation. Dot-dashed green points are the (sub)monolayer structures with (3$\times$3) cell. Orange star and plus are the (7$\times$7) and (8$\times$8) surface monolayer respectively. Red points are HCP films and black are hR2 Bi films on the (3$\times$3) cell. Blue points are HCP Bi films on [3-112] cell.}
\label{fig:enthalpy}
\end{figure}

Fig.~\ref{fig:enthalpy} shows the enthalpy of formation for various surface structures with different thickness. Notice that both the (8$\times$8) and (7$\times$7) surface monolayer structure touch the convex hall which implies they are both energetically stable. The stable (3$\times$3) surface monolayer is equavilent to both 1 layer HCP film and hR2 film with (3$\times$3) cell. Moreover, for higher coverage, hR2 films have much lower energy than the HCP films. For instance, at the same total coverage of 16/9=1.8 Bi/Ni, the total energy the five layer hR2 film is 1.5 eV lower than the four layer HCP film on the bare (3$\times$3) surface. This energy difference is much larger than the thermal energy, $k_BT$=40 meV at 473 K. The hR2 films are thus much more likely to form than the HCP films. The hR2 films favor odd number of layers (surface monolayer + integer bilayers) which is consistent with the experimental observations of 3, 5 and 7 layer films. However, the stability is due to the exotic chemical bonding of Bi rather than QSE.

\begin{figure}[ht]
\includegraphics[scale=0.45,width=0.5\textwidth,clip]{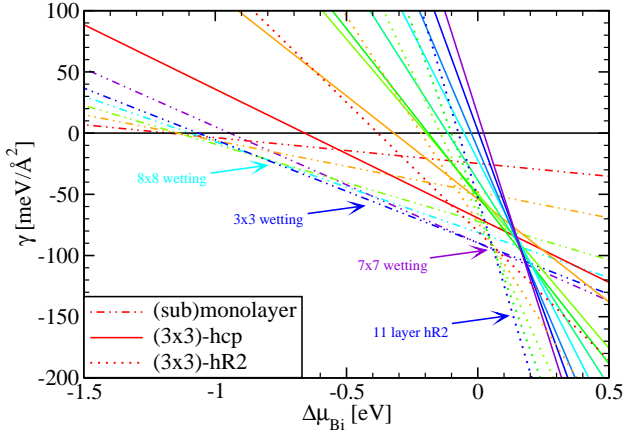}
\caption{(Color online) Surface free energy. Dashed-dotted lines stand for monolayer or less. Solid and dotted lines stand for HCP and hR2 films on (3$\times$3) cells respectively. Different slopes indicate different coverages according to Eq.~\ref{eq:gamma}. Stable sequence is indicated by arrows for monolayer or more.}
\label{fig:sfreeE}
\end{figure}

To further illustrate the stable sequence from equilibrium thermodynamics, we calculate the surface free energy. This quantity is the Legendre transform of the enthalpy of formation (Eq.~\ref{eq:enthalpy}), replacing the surface coverage with relative chemical potential $\Delta\mu_{Bi}$. From equilibrium thermodynamics, the most stable structure at a certain Bi chemical potential $\Delta\mu_{Bi}$ minimizes the surface free energy~\cite{Kitchin08},
\begin{equation} 
\label{eq:gamma}
\gamma =[\Delta H-\Delta\mu_{Bi}N_{Bi}]/A
\end{equation}
where $\Delta H$ is the enthalpy of formation (Eq.~\ref{eq:enthalpy}), $\Delta\mu_{Bi}=\mu_{Bi}-E_{Bi}^{bulk}$ is the Bi relative chemical potential. The results in Fig.~\ref{fig:sfreeE} shows the stable sequence is from bare surface to one atom on the (3$\times$3) cell at $\Delta\mu_{Bi}$=-1.16 eV, to three atoms on the (3$\times$3) cell at -1.11 eV, to (8$\times$8) surface monolayer at -0.83 eV to four atoms ((3$\times$3) surface monolayer) at -0.82 eV, to the (7$\times$7) surface monolayer at .01 eV, and then finally to the infinite height hR2 films at around 0.06 eV (actually we stopped our calculation at 11 layers). Other structures are thermodynamically unstable. However since the growth is kinetically prevented from forming infinite height films, metastable structures with low energies (e.g. finite thickness hR2 films) will appear in actual growth.  However, the HCP-based structures lie systematically above the hR2 ones. By studying the bulk Bi energy with different lateral strains, we confirm that the underlying reason that hR2 films are more favorable than HCP films is that Bi favors a bilayer structure with strong covalent bonding. (see Appendix A1.)

The experimental surface monolayer structure is not clear, as the authors published an erratum~\cite{Bollmann12} regarding their 7-layer film being (8$\times$8) rather than the initially claimed (7$\times$7). We find both (7$\times$7) and (8$\times$8) surface monolayer are stable from our total energy calculation. If the observed surface monolayer is (8$\times$8) rather than (7$\times$7), then a 3 layer hR2 film would fit the experimental deposition rate better than the HCP film would (see Appendix A3.). 

Besides the HCP and hR2 (001) films, we also studied the energy of free standing hR2 (012) films. The 1 and 2 bilayer hR2 (012) films are more stable on Si(111) than hR2 (001) films. However, with the Ni lattice constant rather than Si, the commensurate bilayer (001) film has lower energy by 30 meV/atom, and thus the (012) films are not favorable. This is also consistent with the experimental observation that no pseudocubic structure appears.

\section{Conclusion}

We study the growth of Bi on Ni(111) surface using first principles calculations. The proposed HCP films pucker under relaxation and are energetically and mechanically unstable to adding capping atoms. We find instead that bulk-like (001)-oriented hR2 films above the surface monolayer are more energetically favorable than HCP films. The hR2 films seem to fit with experimental observations (LEED pattern and coverage) equally well as HCP. If our model is correct, growth on Ni(111) might provide a useful synthesis of uniform hR2 bilayers, which have been shown to act as two-dimensional topological insulators~\cite{Hirahara11,Yang12}. We hope our theoretical work can trigger more interesting work, both theoretical and experimental in this subject.

Besides the surface growth, phenomena of Bi at Ni interfaces also attract attention recently. Liquid Bi penetrates and segregates at Ni grain boundaries forming bilayer structures~\cite{Luo11} in a stable grain boundary phase called a complexion~\cite{Harmer11,Dillon07}. These bilayer interfacial structures can possibly explain the long standing puzzle of the liquid metal embrittlement. However, the underlying mechanisms of bilayer segregation and their relation with embrittlement have not been revealed at the quantum level. Our study of Bi on Ni surfaces serves as a precursor to this interfacial study. In particular, we note that a pair of surface monolayer films, one on each surface at a grain boundary, provides an attractive model for the observed Bi bilayers.

\section{Acknowledgment}
The authors thank Zhiyang Yu, Di Xiao and Randall Feenstra for helpful discussion. Financial support from the ONR-MURI under the grant NO. N00014-11-1-0678 is gratefully acknowledged.

\bibliographystyle{apsrev}
\bibliography{BiNi}

\section{Appendix}

\subsection{A1. Lateral strain}
\begin{figure}[t]
\includegraphics[scale=0.45,width=0.5\textwidth,clip]{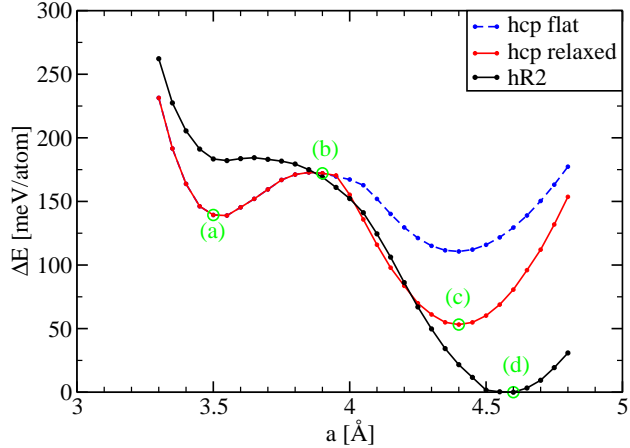}
\caption{(Color online) Bi bulk energy of HCP and hR2 structures. Black is the relaxed hR2 structure. Red is the relaxed HCP structure. Blue is the evenly spaced HCP structure. Points (a),(b),(c),(d) correspond to the structures whose energies are plotted in Fig.~\ref{fig:film}.}
\label{fig:bulk}
\end{figure}

The proposed Bi HCP films have in-plane lattice constant $a=3.7$~\AA~(3$\times$3) and 3.8~\AA~([3-112]), while the hR2 films commensurate to Ni(111) has $a=4.3$~\AA. To explain why the hR2 films are more favorable than the proposed HCP films we investigate the bonding character of Bi by calculating bulk Bi energies with different $a$. We adopt bulk cells with a (1$\times$1) unit cell in the $xy$ plane and 6 atomic layers in the $z$ direction. The $a$'s are fixed while lattice constants $c$'s are fully relaxed. For $a$ less than 3.9~\AA~, both the relaxed HCP and hR2 structures are evenly spaced. For larger $a$ values, Bi in both structures pairs up to bilayer structures. Fig.~\ref{fig:bulk} shows Bi HCP and hR2 bulk energies with fixed $a$. Clearly, the paired hR2 bilayer structures with $a\approx4.6$~\AA~are more favorable than the proposed HCP structures with $a\approx3.5$~\AA. By paring up, the Bi chemical bonds change from metallic to covalent in nature. This strongly affects the periods and amplitudes of surface energy oscillation of free standing Bi films. 

\subsection{A2. Quantum size effect}

In thin metallic films, electrons are confined in the verticle direction. At low temperature, the energies of the confined electrons varying with film thickness governs the relative stability of the films. This leads to ``electronic growth''~\cite{Zhang98}, a type of QSE. Based on the usual quantization rule, the energy oscillation period in a free electron model is half of the Fermi wavelength~\cite{Miller88}. In a solid, taking account of the band structure, the actual energy oscillation is the superposition of different oscillations at high symmetry points in the surface Brillouin zone~\cite{Wei07}. For HCP metal, the (001) electron confinement energy can be written as,
\begin{equation}
E(N)=A_{\bar{\Gamma}}\sin{(2k_{\bar{\Gamma}}Nd+\phi_{\bar{\Gamma}})}+A_{\bar{M}}\sin{(2k_{\bar{M}}Nd+\phi_{\bar{M}})}+A_{\bar{K}}\sin{(2k_{\bar{K}}Nd+\phi_{\bar{K}})}
\end{equation}
where $\bar{\Gamma}$, $\bar{M}$ and $\bar{K}$ are three high symmetry points in the surface Brillouin zone, $k$'s are the Fermi wave vectors, $A$'s characterize the importance of those three points, $\phi$'s are the phase shift of three oscillations, $N$ is the number of layers and $d$ is layer spacing. The resulting energy is the superposition of those three oscillations. We compare the band structure prediction and the total energy oscillation of Bi films. 

\begin{figure}[b]
\includegraphics[scale=0.9,width=1.0\textwidth,clip]{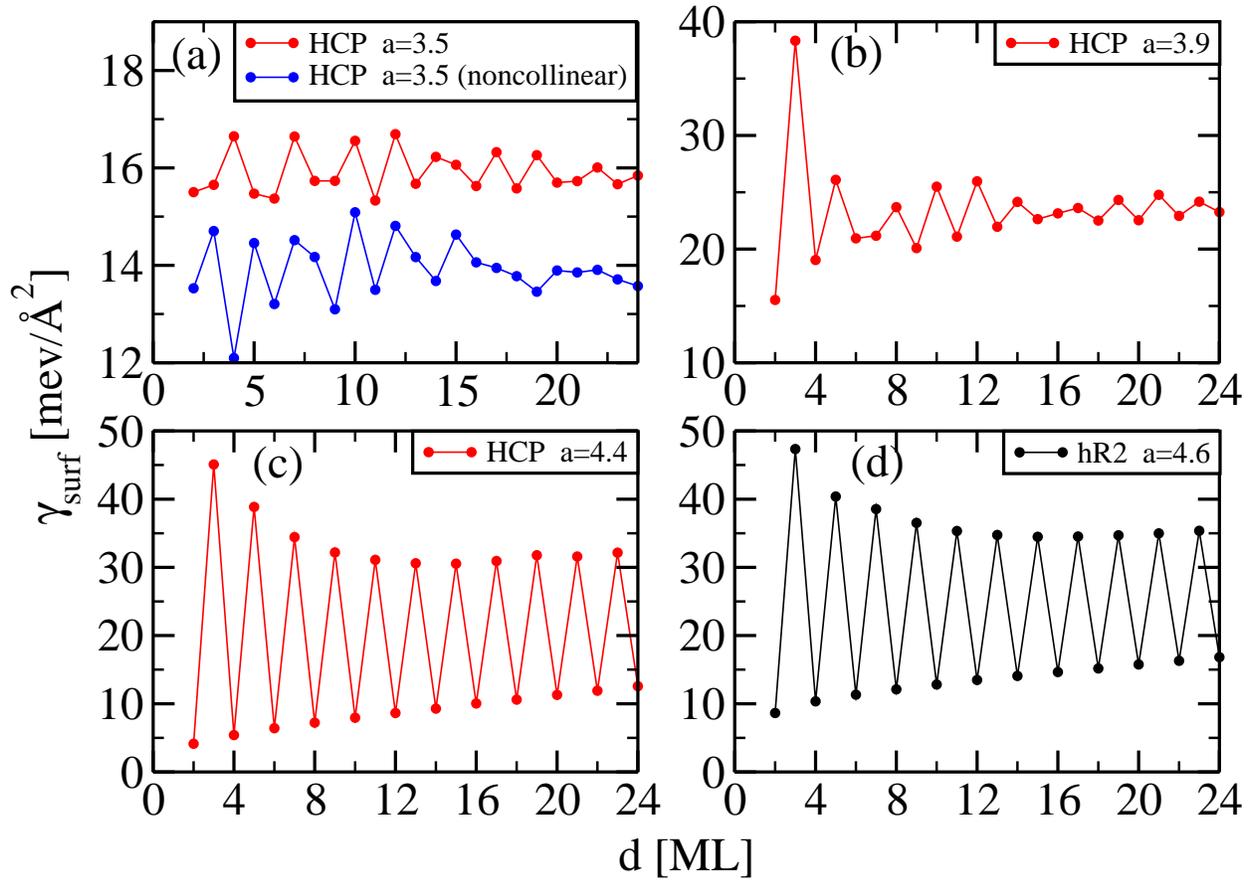}
\caption{(Color online) Surface energies of free standing Bi films with different in-plane lattice constants $a$. Black and red curves are using collinear calculation. Blue using noncollinear.}
\label{fig:film}
\end{figure}

Fig.~\ref{fig:film} shows the surface energy oscillation with different $a$ values. Here the surface energy is defined as,
\begin{equation}
\gamma_{surf}=\frac{1}{2A}[E_{tot}-E_{Bi}N_{Bi}]
\end{equation}
where $E_{tot}$ is the energy of the film, $N_{Bi}$ is the number of Bi atoms in the film, $A$ is the surface area and $E_{Bi}$ is the linear part of the total energies of the film structures as in Eq.~\ref{eq:surfNi}. Shown in Fig.~\ref{fig:film}(a), for HCP films with $a=3.5$~\AA, the oscillation is complex in both period and amplitude reflecting the superposition of multiple periodicities. Fourier analysis of $\gamma_{surf}$ yields a period of around 3 layers. From our band structure calculation, the bulk Fermi wave vectors at $\bar{\Gamma}$, $\bar{M}$ and $\bar{K}$ are, respectively, $0.59\pi/d$, $0.33\pi/d$ and $0.13\pi/d$. Averaging over these three frequencies, the resulting oscillation period is 2.9 layers, which agrees well with the total energy calculation $\gamma_{surf}$. This QSE due to electron confinement does govern the energy oscillation of the free-standing HCP film with $a=3.5$~\AA. However, the energy oscillation amplitude is much smaller than the energy oscillation of Bi HCP films on Ni(111), implying that quantum size effect is not the dominant factor in determining the stability of Bi HCP film on Ni(111). Also unlike Bi on Ni(111), SOC does alter the relative stability of free standing HCP films with $a=3.5$~\AA.

Fig.~\ref{fig:film}(b) shows the surface energies of HCP films with a equal to 3.9~\AA~where the bulk Bi atoms start to pair up. The energy oscillation period is close to 2 layers which is different from the band structure prediction of 3.2 layers. Fig.~\ref{fig:film}(c) and \ref{fig:film}(d) show the energies of HCP films with a equal to 4.4~\AA~and hR2 films with a equal to 4.6~\AA~respectively. In both cases, the energies show bilayer oscillation with much larger amplitude than the oscillation due to this QSE. We conclude that the oscillations of $\gamma_{surf}$ are not due to this QSE for $a>3.9$~\AA~instead it is due to covalent bonding into bilayers which is a different type of QSE.

\subsection{A3. Deposition rate}
We analyze Fig. 1 in the support material of~\cite{Bollmann11} to estimate the coverage of the 3$\times$3 film. As the deposition rate is constant, the total number of Bi atoms doubles from deposition time of 6000s to 12000s. The surface monolayer has fractional area of around 90\% at 6000s and around 20\% at 12000s. Since the fractional area of [3-112] is small, we just ignore the small coverage difference between it and the (3$\times$3) film in our estimation. We then get the equation,
\begin{equation}
0.9\times\theta_{mono}+0.1\times\theta_{3\times 3}=\frac{1}{2}(0.2\times\theta_{mono}+0.8\times\theta_{3\times 3}) 
\end{equation}
where $\theta_{mono}$ and $\theta_{3\times 3}$ are the coverage of surface monolayer and 3$\times$3 film respectively. We then get,
\begin{equation}
\theta_{3\times 3}=2.7\times\theta_{mono}
\end{equation}
The resulting coverage of the (3$\times$3) film is $\theta_{3\times3}=1.4$ Bi/Ni if the surface monolayer is (7$\times$7) with $\theta_{mono}=25/49=0.56$ and $\theta_{3\times3}=1.1$ if the surface monolayer is (8$\times$8) with $\theta_{mono}=25/64=0.39$. In comparison, the 3 layer (3$\times$3) HCP film has coverage $\theta_{3\times3}=12/9=1.3$ and the 3 layer (3$\times$3) hR2 film has coverage of $\theta_{3\times3}=10/9=1.1$. The hR2 film thus fits with the deposition rate better than the HCP film if the surface monolayer is (8$\times$8), while the HCP film fits better if the surface monolayer is (7$\times$7).

\end{document}